# Towards a Holistic CAD Platform for Nanotechnologies

E. Kolonis, M. Nicolaidis

TIMA Laboratory

**ABSTRACT**

Silicon-based CMOS technologies are predicted to reach their ultimate limits by the middle of the next decade. Research on nanotechnologies is actively conducted in a world-wide effort to develop new technologies able to maintain the Moore's law. They promise revolutionizing the computing systems by integrating tremendous numbers of devices at low cost. These trends will provide new computing opportunities, have a profound impact on the architectures of computing systems, and require a new paradigm of CAD. The paper presents a work in progress on this direction. It is aimed at fitting requirements and constraints of nanotechnologies, in an effort to achieve efficient use of the huge computing power promised by them. To move towards this goal we are developing CAD tools able to exploit efficiently the huge computing capabilities promised by nanotechnologies in the domain of simulation of complex systems composed by huge numbers of relatively simple elements.

## 1. INTRODUCTION

Shrinking of silicon-based technologies will reach its ultimate limits in about one decade. The reasons are not only technical (leakage current, signal integrity, energy consumption, circuit heating…), but also economic (excessive cost of fabrication lines) [1]. Single-electron transistors, quantum cellular automata, nano-tubes, molecular logic devices, nano-crystals, are some of the candidate alternative technologies [2]-[9]. In their majority the related fabrication processes emerging in various research centers are bottom-up, taking advantage of the self-assembling properties of atoms to form molecules and crystals. They should allow the fabrication of very complex circuits at low cost and low energy consumption [10]-[12]. Some of the characteristics of such circuits will be: an extraordinary complexity, a high regularity, and a low fabrication yield. One consequence is that these processes can not produce highly sophisticated circuits with a high morphological differentiation, similar to MOS circuits. The fundamental reason yielding this kind of structures is that a self-assembling process consists on replicating a few "simple" basic modules a large number of times to form a regular structure (hereafter referred as nano-network). Of course, the basic modules may gain a certain complexity as these techniques gain on sophistication.

By integrating tremendous numbers of devices at low fabrication cost, the self-assembling processes should allow revolutionizing the computing systems. On the other hand, due to their intrinsic characteristics and constraints, these trends will have a profound impact on future architectures, requiring possibly shifting into a non von Neumann, highly parallel, reconfigurable computing paradigm for their efficient exploitation [12]-[15].

A pioneering work [16] carried on at Hewlett-Packard Laboratories illustrated an approach able to tackle some of the constraints related to these technologies. They developed the TERAMAC computer, a multiprocessor built by interconnecting a large number of FPGAs. They used defective FPGAs and error-prone technologies to interconnect them, as an illustration of hardware platforms composed of a large number of identical, but error prone modules. The multiprocessor was created by programming conventional processor architectures into the regular structures of the FPGAs. Prior to this mapping, test and diagnosis procedures were applied to identify the defective elements of the system. Then, the mapping was done in a manner that only fault-free resources were employed. This work illustrates the feasibility of implementing complex systems by programming a regular network composed of a large number of simple and identical modules, affected by high defect densities. It provides an additional motivation for using regular, FPGA-like, structures in nanotechnologies, since achieving fault tolerance for the expected defect densities would be much more difficult and costly for other circuit structures. This fault tolerance approach is very general, and could be used with various reconfigurable architectures. But, as discussed in chapter 2, some other aspects must also be addressed for exploiting efficiently the promise of nanotechnologies on producing low-cost extraordinary complex systems.

In this paper we describe a CAD platform that would allow tomorrow's engineers to exploit efficiently the extraordinary computing power promised by nanotechnologies, in order to model and simulate complex natural or artificial systems composed of huge numbers of simple elements. This platform is said holistic, because, it includes tools enabling a high-level description and simulation of the target application (e.g. a an ecosystem, a set of interacting particles, …), down to tools enabling efficient circuit-level implementation.

## 2. A HOLISTIC CAD PLATFORM FOR NANOTECHNOLOGIES

The present work targets CAD tools allowing efficient use of the hardware resources promised by nanotechnologies. It is motivated by the following converging factors:

a- The availability of hardware platforms comprising huge amounts of identical modules interconnected into a regular network (the promise of nanotechnologies).

b- The flexibility of these platforms, which allow implementing various computing architecture by programming them.

c- The profound interest for science and technology to explore the dynamics of complex natural and artificial systems composed of huge numbers of identical elements.

Point c reflects the fact that natural systems are often composed of huge numbers of simple elements. The ability for simulating accurately natural systems composed of huge amounts of particles (e.g. atmosphere and oceans, chemical reactions, nuclear reactions, solar system formation, galactic interactions, …) and biological systems (molecular interactions in the cell, cellular interactions in an organism, cellular differentiation in the embryo, neural systems, interactions between organisms in an ecosystem, ….), is a fundamental requirement for the computing systems of the future. In addition to natural systems, experimenting invented systems, composed of large numbers of simple modules, is convenient for studying the impact of the laws governing the basic modules on the global dynamics of this kind of systems.

The availability of powerful hardware platforms, as promised in point a, will allow to increase drastically the accuracy of our simulations for both, the natural and artificial systems discussed above, and push drastically our understanding of such systems. By using a "traditional" approach, as the one adopted in the TERAMAC project, we can implement such applications in two steps:

- First, program the nano-network to map a general purpose multiprocessor architecture.
- Then, program this architecture to simulate the target regular system.

But this approach results on severe resource waste, since:

- Parallel computers based on conventional architectures make poor usage of their theoretical computational power.

- General-purpose architectures waste the hardware resources with respect to architectures dedicated to target applications.

A drastic gain in efficiency could be obtained if we program the nano-networks to map directly the target application. In other words, we will map the basic module of the target system on a set of modules of the nano-network. This mapping will be repeated to program as many modules of the target system as allowed by the complexity of the nano-network. This direct implementation of the complex system could be done efficiently if the nano-network architecture supports certain reconfigurable structures and the mapping is done in a manner that establishes the required communication between the modules of the target system. Furthermore, to make the approach practical, a dedicated CAD platform has to be developed. Because this platform should enable the direct implementation of the final application in the nano-network, it should include high-level tools enabling easy software-level description and simulation of the target application down to tools enabling its efficient implementation at circuit-level, hence the term holistic. This is not the case of today's EDA platforms as the current practice is to develop general purpose computers on which the final applications are programmed, or, in some particular cases, implement specific circuits accelerating power-consuming computations for a class of applications. But, in the later case again, the final application is implemented by software. Therefore the proposed CAD platform will comprise:

**A High-level Modeling and Simulation tool (HLMS):** it enables easy generation of a software description (model) of the target complex system and its exploration through simulation on conventional computers.

**A Nano-network Architecture Fit tool (NAF).** It transforms the description of the complex system generated by the HLMS tool into a description that fits the constraints of the nano-network architecture.

**A Circuit Synthesis tool (CS).** It maps the description generated by the NAF tool onto the configurable architecture of the nano-network.

These tools together with the constraints imposed by the nano-netwok architectures are discussed in the next sections.

## 3. HIGH-LEVEL MODELING AND SIMULATION TOOL (HLMS)

This tool allows easy description of the target complex system by means of an interactive menu. Then, it generates a software description (model) of the complex system. Based on this description the tool can perform preliminary simulation on conventional general-purpose computers. Thus, before engaging the complex task of implementing the complex system in the nano-network, the user can simulate on conventional computers a reduced version of the system to validate his/her various choices concerning the system parameters and the evolution laws of the system entities. More importantly, the HLMS tool is mandatory for generating the software description of the target application which is the starting point for the other components of our synthesis platform. Consequently, our efforts are first concentrated on the development of this tool.

The HLMS tool and its experimentation over two families of complex systems are presented below. A more detailed description will be provided in [25].

The complex systems considered are systems composed of a large number of interacting entities, in which a rich behavior emerges as a consequence of the large number of entities and of their interactions rather than as a consequence of the complexity of these entities. The entities composing the complex system can be identical or can belong to a small set of entity types. Each entity type will be described by:

- A set of state variables
- A set of communication signals between modules
- A set of functions that determine the evolution of the state variables of each type of module.

The evolution of the state of the elementary entities and, thus, of the global system is produced in the following manner. At each simulation cycle, the evolution functions use the present state of the state variables and the incoming communication signals to produce the state of the entity's variables for the next simulation cycle (see Figure 1).

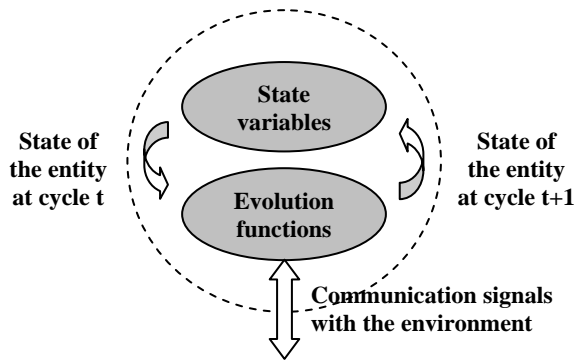

Figure 1: elementary entity

Thanks to a graphical interface, the user can specify the fundamental structure of his target complex system. This is done by specifying the set of internal variables of each entity type of the system, the inputs received from and the outputs send to the entities with which the specified entity interacts, and the evolution rules that determine the next state of each internal variable and of each output of this entity as a function of its present state and of its inputs. To make the system intelligible, this interface allows also specifying some visualization parameters (e.g. associate different colors that reflect the type or the state of an entity or specify graphical representations of different parameters). The graphical interface allows modifying the visualization and the other parameters of the system even during simulation

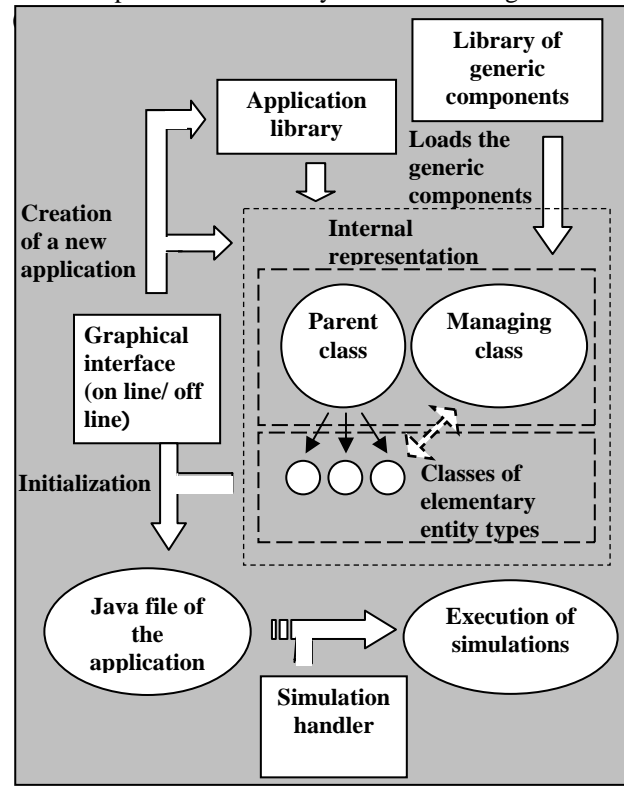

Figure 2: Structure of High-level Modeling and Simulation tool

To implement the High-level Modeling and Simulation tool we adopted an object oriented approach based on the Java programming language. The structure of the tool is shown in Figure 2. It consists of a graphical interface, an application library, a generic component library and a simulation handler module. The tool also includes a central program not shown in the figure. As said earlier, the user specifies his/her target system by means of the graphical interface. The user can save this description in the application library for future use. Furthermore, the central program of the tool uses this description as well as the description of some generic components from the library of generic components to generate an internal representation of the application. This representation contains several classes:

- **Classes of elementary entity types**. For example, in the case of an ecosystem, these entities can correspond to the types of cells of an organism (stomach cells, isolation cells …)
- **Parent class** from which the classes of elementary entity types can inherit the variables and functions that are common for all. For instance, in the case of an ecosystem, this class can contain variables such as position and heat and the functions of movement and heat diffusion that are common to all the cells of an organism.
- **A managing class** that manages the evolution of the state of the elementary entities and can, during the simulation, create new entities if this is allowed by the laws governing the system.

The user can employ the graphical interface to initialise the internal representation of the application. Then, the central program of the tool generates a file called "java file of the application". Finally the simulation handler uses this file to execute simulations.

The adopted structure and implementation are generic and allow handling various kinds of systems composed of large numbers of interacting entities. To validate its versatility, we used the tool to experiment two quite distinct families of complex systems. The first family consists on artificial ecosystems and the second consists on artificial "universes" composed of a set of interacting particles governed by laws that determine the geometry of space-time (relativistic space-time in the example considered here) [24].

In addition to the generic modules of the High-level Modeling and Simulation tool, it is useful to implement specific modules for each family of complex systems to easy the user's tasks for creating, experimenting and analyzing systems belonging to this family.

### 4. ARTIFICIAL ECOSYSTEMS

Our first experiments concern artificial ecosystems consisting on an artificial environment (the space of the ecosystem, the milieu which may have properties like temperature and resistance to the motion of objects, food/energy sources, …) in which evolve virtual multi-cellular organisms with properties inspired from the real world, such as cell replication and cell differentiation leading to an adult organism from a single cell (ontogeny), movement, food capturing, sexual reproduction, adaptation to the environment, and phylogeny through mutation, crossing-over, and natural selection. Similarly to the natural organisms, the cells of an artificial organism all share the same genetic information (coded into "DNA" variables) and obey to the similar elementary functions as those described by molecular biology.

From the computational point of view each module (a cell of an organism), is viewed as an automaton. It has a certain number of inputs and outputs, a certain number of internal variables, and a function associated to each variable. This function computes the next state of the variable from the present state of all the internal variables and the present values of the cell inputs. A similar function is also associated to each cell output. The graphical interface of the High-level Modeling and Simulation tool is used to specify these variables and their associated functions and subsequently the other parameters of the ecosystem. Furthermore, in order to easy the users' tasks we implemented some functional modules. In particular:

**An ontogeny module:** this module assists the user to code in the DNA variables the genetic information that guide the ontogeny process of an organism.

**A module of reproduction and creation of organisms:** this module manages the production of the new DNA of a children organism by the processes of crossing-over and genetic mutations.

**A module of "natural selection":** it allows the user to easily experiment the "natural selection process" by selecting and eventually modifying on-line (i.e. during the simulation). This is possible because the user graphical interface (figure 2) allows modifying the parameters of the system even during the execution of a simulation.

Thanks to these modules, the user can easily create and experiment a large variety of artificial ecosystems as illustrated below.

### 4.1 Spontaneous creation of artificial organisms

In this experiment, we consider an environment with the potential of appearance of DNA variables with arbitrary values associated to various functions contained in a library (library of DNA functions). In the beginning of the simulation, we observed the appearance of DNA combinations that led to the formation of organisms in an arbitrary way. In the majority of the cases, the spontaneously created organisms lacked in complexity and ingenuity and survive for short time. As the simulation runs for longer time, appear by chance some organisms which survive longer because their DNA combinations are more efficient and they begun to populate the ecosystem. In the long run, we observe an evolution towards forms which gain in complexity and are better adapted to the environmental conditions. Thus, we observed the emergence of organisms that could move and feed (from food sources randomly distributed in space and time), reproduce, capture another organism (prey) in order to feed, ...

### 4.2 Optimization of existing functions

These experiments concern the optimisation of existing organs to adapt to the environmental conditions. We consider as example the influence that has on the size of the motion organs the factor of resistance ($Fr$) that the milieu opposes to the motion of objects. We observe that for higher values of $Fr$ the size of the motion organs increases and the organism spends more energy for movement. This is necessary in order to capture enough food to cover its other energy needs (for basic metabolism and temperature maintenance). We also observe an increase of the population when $Fr$ decreases. Obviously, in this case, the organisms can spend more energy for reproduction (the energy saved from the movement). Both results illustrate a good adaptability of the artificial organisms to the evolution of the environment. Numerous experiments on other genes also shown good adaptability.

### 4.3 Acquisition of new functions

These experiments illustrate another ability of the tool to create pertinent artificial ecosystems: the evolution towards more efficient organisms by acquisition of new functions. The library of functions contains among others an olfaction function (sensitivity to the concentration of smell substances (cells) released by the food sources) and a function of motion-direction control that reacts to the state of other organs. The exact manner the motion control function reacts to this state is determined by DNA values associated to a large number of functions. Initially the organisms of the ecosystem do not posses olfactory and motion-control organs and move randomly. Thus, they discover food sources only by chance. After a certain time we observed the emergence of organisms possessing olfaction organs, and other processing motion-control organs. Later, new mutations created organisms that possess both functions but their motion control functions are not effective. Later emerged organisms that possess several olfactory organs placed at different parts of their body as well as motion-control organs that direct the motion towards the direction of those olfactory organs that sense higher concentrations of smell substances. In environments with sparse food sources, these organisms replace quickly all other organisms as they grasp almost all food. This example illustrates an evolution-based learning that codes intelligent behavior in simple genetically-programmed functions.

A next step in our experiments will consist on exploring the emergence of ant-kind collaborative societies based on genetically-programmed functions. A further step will consist on introducing neural-type functions. The facility offered by the tool to create and experiment an "infinity" of artificial ecosystems by expending the library of functions and by playing with the environmental and other parameters allows exploring a vast universe of artificial ecosystems.

## 5. SYSTEMS OF PARTICLES AND EMERGENCE OF RELATIVISTIC SPACE-TIME

To illustrate the versatility of the High-level Modeling and Simulation tool we have considered a second family of complex systems referred hereafter as artificial "universes". They consist on large numbers of elementary particles whose interactions obey a set of laws specified by the user. The generic modules of the HLMS tool allow easy creation and simulation of such systems. However, our goal was to go one step further and study global properties of such systems and in particular the structure of their space-time. Such experiments are useful to support new interpretations of modern physics. Since several millenniums, we consider that our world is composed of a set of objects evolving in space with the progress of time. The evolution of objects and some of their properties should therefore conform to the structure of space and time. In particular, in special relativity, the structure of space-time is described by the Lorentz transformations and imposes the modification of the length of physical objects (length contraction), the reduction of the pace of evolution of the physical processes (time dilatation), and the loose of synchronization of distant events when observed from different inertial frames. As a consequence of this vision, two theories are needed to describe the laws governing the universe, one to describe the structure of space-time and another to describe the behavior of elementary particles (the elementary units that form the physical objects and the physical processes).

Inspired from the theory of computing systems we can view the universe as a system composed of elementary computing modules (elementary particles), which determine their next state as a function of their present state and the state of the modules with which they interact. In this vision, the form of the space at any instance of the system's evolution is determined by the values of the position variables of the elementary modules. The structure of time is determined by the relationships of the paces of evolution of the different processes taking place in the system. Thus, this vision requires only a theory describing the laws governing the evolution of elementary particles, since these laws determine the evolution of the system, which determines on its turn the structure of space and time.

As shown in [24], the space-time emerging in such a system obeys Lorentz transformations if and only if the laws of interaction of elementary particles obey a certain condition referred as Relativistic Constraint of Accelerations (RCA). The central idea of this condition is that the intensity of the interactions changes accordingly to the speed of the particles. As a consequence, the distances of mutual equilibrium of the particles forming a moving object are modified (contraction of lengths), the pace of the evolution of a group of particles (a process) is modified (dilatation of time), and the synchronism between distant events is altered.

The aim of the experiments presented in this section is to illustrate this vision. This is done by simulating artificial "universes" in which the laws of interactions between elementary particles obey the RCA condition and by showing that the results of measurements performed in different inertial frames obey the Lorentz transformations. Thanks to the HLMS tool, we can easily create and simulate artificial "universes" governed by interaction laws obeying the RCA condition. However, we also need new modules that allow measuring object lengths and process durations in various inertial frames. As a matter of fact, we implemented some new modules facilitating performing such measurements: They include:
- A module for creating a length reference (length unit) composed of a set of particles forming a rigid object. This object is created in any inertial frame and at any orientation specified by the user.
- A module for creating a time reference (clock) composed of a set of particles that produce a periodic process of constant period. This clock is created at any positions and any inertial frames specified by the user.
- A module for synchronizing clocks placed at any inertial frame and any locations specified by the user.

We used the tool to create and experiment artificial "universes" whose laws of interactions obey the RCA condition. We experimented laws which, for particles at rest (position variable is constant), take various forms such as $r^{-2}$ (r being the distance between the interacting particles), or more "strange" forms such as $P(r)/Q(r)$ where $P(r)$ and $Q(r)$ are arbitrary polynomials of r, or $e^{Q(r)}$. For particles moving with arbitrary speeds these laws were adapted to conform the RCA condition. In all cases, as expected from the theoretical results in [24], all measurements were in conformity with Lorentz transformations.

## 6. ARCHITECTURAL SOLUTIONS AND NANO-NETWORK ARCHITECTURE FIT TOOL (NAF)

The HLMS tool generates a software description of the target application executable in conventional computers. The implementation and execution of the target application in a nano-network is conditioned by the particular architecture of the nano-network. The aim of the NAF tool is to transform the description generated by the HLMS tool into a description that fits the constraints of this architecture (figure 3). Furthermore, the NAF tool will allow comparing the descriptions of various implementation options of the target system, in order to determine the most efficient ones. The description generated by the NAF tool includes two parts, the description of the function of the basic modules of the complex systems and the description of the communication of these modules. The former will be identical to the one generated by the HLMS tool. The generation of the later has to take into account the communication resources of the nano-network and the strategies that could best solve the communication strangle, which, by the way, is the most challenging task in highly parallel systems. Thus, determining relevant nano-network architectures and communication strategies is a task mandatory for implementing an efficient NAF tool. These aspects are discussed below.

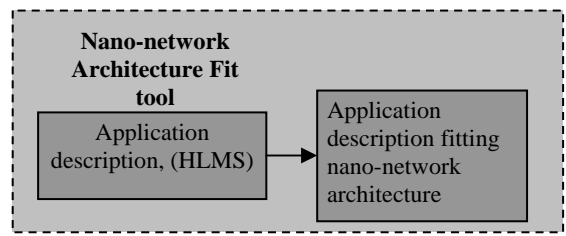

Figure 3: The architectural exploration tool

**Once** the functions of the modules (inputs, outputs, internal variables and evolution rules) have been determined, the system can be programmed in a nano-network, consisting on a programmable (FPGA-like) hardware. However, efficient exploitation of the hardware resources depends on the architecture of this hardware and the way a target application is mapped on it. For applications where the modules are not mobile and their function is stable in time, we can map the function of the modules on this hardware, similarly to the mapping of traditional hardware on FPGAs. On the other hand, for applications using mobile modules, which eventually have changing functions, managing the interactions between modules is very challenging. Let us analyze the challenging cases:

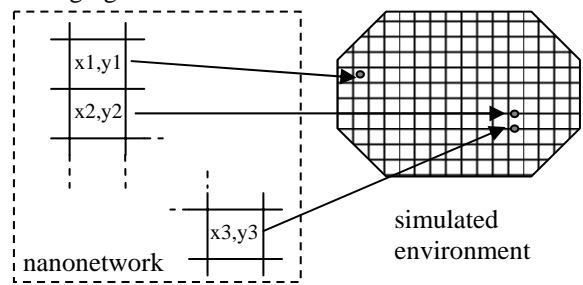

Figure 4: The first option for supporting applications with moving modules

To support the movement of modules we can use three options:

**a**- The first option uses a *position variable* to code the position of each module in the environment. The module moves in the environment by changing the value of this variable, but its position in the nano-network remains stable. Thus, modules located in physically distant locations within the nano-network may be neighbor in the environment and have to interact (see figure 4, modules [x2, y2] and [x3, y3]). For performing the interaction between such modules the nano-network should either support interconnections of exponential complexity (in order to be able to connect at any time any pair of locations), or it will need to spend a number of computation cycles proportional to the size of the network (in order to transfer the interaction information using local only connections). These solutions affect severely either the size of the hardware required to map a given application or the speed of the computation. To avoid these problems, we must implement the application in a manner that only local interactions are used, as described below.

**b-** The second option does not use a position variable for each module. Instead, the position of each module within the environment is identical to its physical position in the nano-network. In this case, when a module is moving within the environment its implementation must also move within the nano-network. For applications where all modules are identical, all the areas of the nano-networks implement the same function. Thus, module movement requires passing the values of the variables of a module to an unoccupied neighbor location of the nano-network, or exchanging the values of the variables of two modules occupying neighbor locations in the nano-network. This can be done easily with conventional FPGA-like architectures (non-reprogrammable or reprogrammable). However, if the application uses modules of several types, then, moving a module requires passing to a neighbor location not only the values of its variables but also its function, resulting on intense reconfiguration process. In conventional reprogrammable FPGA architectures this task is possible but very time consuming, since the reconfiguration information has to be transferred to an external system that controls the programming of the network, and from this system to be written to the configuration memory for reprogramming the function of the considered locations. Doing so for a large number of moving modules will affect dramatically the performance of the system. Thus, conventional FPGA-like architectures are not suitable. Instead, the architecture of the nano-network should provide connections for exchanging between neighbor cells the values programming their function. This solution enables also efficient hardware use in applications where the function of modules can vary, as a result of their evolution.

**c-** However, in applications where the different modules occupy a small amount of the space in which they evolve, using as position of a module in the simulated space its physical position in the nano-network, as suggested in point b, will waste a large amount of hardware resources. This will happen because all unoccupied positions of the simulated space will correspond to unused locations of the nano-network. To cope with, we propose a solution that merges the above cases a and b. With this solution, the position of each module is determined by an internal variable as in case a. In order to move, the module increases or decreases its coordinates (e.g. the x and/or y component of the position variable in a two-dimensional space). However, to solve the communication problem between modules we impose some constraints that maintain a spatial coherence between the positions of the modules in the simulated space and their positions in the nano-network. A module occupying position (a, b) in the nano-network will be mentioned as module (a, b). Its neighbors in the nano-network occupy positions (a-1, b-1), (a-1, b), (a-1, b+1), (a, b-1), (a, b+1), (a+1, b-1), (a+1, b), (a+1, b+1). We impose the following *spatial constraints* between the position (a, b) in the nano-network and the position variable (x, y) in the environment: The x component of the position variable of module (a, b) never takes a value lower than the x component of the position variables of modules (a-1, b-1), (a-1, b), (a-1, b+1). Also, this component never takes a value higher than the x component of the position variables of modules (a+1, b-1), (a+1, b), (a+1, b+1). The y component of the position variable of module (a, b) never takes a value lower than the y component of the position variables of modules (a-1, b-1), (a, b-1), (a+1, b-1). Also, this component never takes a value higher than the y component of the position variables of modules (a-1, b+1), (a, b+1), (a+1, b+1). With these constraints, the modules surrounding the module (a, b) in the simulated space are the same as in the nano-network (i.e. (a-1, b-1), (a-1, b), (a-1, b+1), (a, b-1), (a, b+1), (a+1, b-1), (a+1, b), (a+1, b+1)). Thus, (a, b) will only interact with its eight neighbors in the nano-network, requiring only local communications. While this solution maintains local communications as in point b, it does not require using as many modules as the locations of the simulated space. In fact, since, a module can move by modifying its position variable, we do not associate a location of the nano-network to each unoccupied position of the simulated space, avoiding the waste of the hardware resources. It is therefore suitable for applications where the modules occupy a small amount of the simulated space (e.g. in an artificial ecosystem living organisms occupy a small amount of the environment). But, as a counterpart, we need to maintain the spatial constraints announced earlier. For that, each module receives from its eight neighbors in the nano-network the value of their position variable. When the new values of the position variables of two modules do not conform to the spatial constraints, the mapping of the modules in the nano-network is modified to satisfy these constraints (an example is given in figure 5).

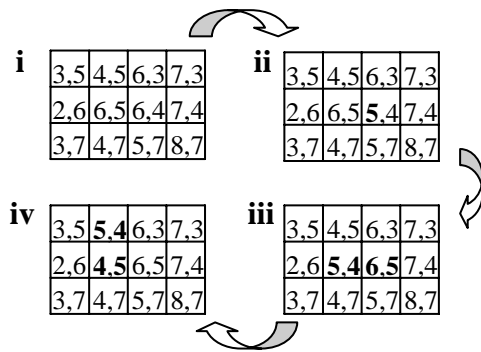

Figure 5: Example maintaining the spatial constraints in case c.
i-   the spatial constraints are respected
ii-  a module moves in the simulated environment and destroys the conformity to the spatial constraints
iii- two modules exchange their positions in the nano-network but the balance is not yet restored
iv-  another change is made in the nano-network and the spatial constraints are restored

The remapping is done locally, by using a nano-network architecture allowing local reprogramming as described in point b. In some situations, this remapping may involve all the modules of a certain area of the nano-network, and may take several cycles since the remapping of some cells may involve the remapping of new cells. But excepting very special situations, the remapping will not involve large areas of the nano-network, since it is not propagated through the empty space surrounding a group of neighbor modules (e.g. the cells of a living organism or of a group of living organisms). During the remapping of the modules, when a location of the nano-network is reprogrammed, it cannot compute and send valid outputs to its neighbors. Thus, it activates a signal to inform them. In response, the neighbors do not perform new computations until this signal is deactivated. In addition, at the next clock cycle they activate a similar signal to inform their neighbors, which interrupt computation on their turn, and so on. A wave of interruptions propagates through the nano-network, but it is stopped when the empty space surrounding the cells of an organism or of a group of organisms is reached. Thus, these interruptions remain local and do not affect the operation of the rest of the network.

The above discussion determines three architectures:
1. For applications using modules deprived of movement, conventional FPGA-like architectures (reprogrammable or non-reprogrammable) can be used.
2. For applications using a single module-type provided with movement, conventional reprogrammable FPGA-like architectures can be used. The position of each module will be determined by a position variable.
3. For applications using multiple types of modules which possess movement capabilities, we will employ architectures allowing local transfer of the function of one location of the nano-network to a neighbor location. Two different approach will be used to determine the position of each module in the simulated space:

3a- if the simulated space is of similar size with the space of the nano-network, this position will be identical to its physical position in the nano-network.

3b- if the simulated space is much larger than the space of the nano-network, this position will be determined by a position variable. In addition, spatial constraints will be imposed to the position variable, as described earlier in point c.

These approaches may solve the communication problem for various complex systems, which is known to be the most challenging problem in highly parallel architectures. This problem is exacerbated by the total parallelism of the approach considered here. However, the criteria for determining the best architecture for an application may not be clear (consider for instance the criteria for selecting approach 3a or 3b). One of the goals of the NAF tool is to enable making such choices. Thus, this tool should be able for each of these choices to create a description fitting the nano-network architecture and then to evaluate its efficiency. This tool is in many aspects similar to the High-level Modeling and Simulation tool as it generates descriptions similar to the ones generated by it. The difference is that it introduces in the descriptions of the system modules and of their communications the constraints of the selected implementation. For instance, in case 3b, it introduces the spatial constraints imposed to the position variable of each module. The next step of this work is to implement the NAF tool. This step consists on modifying the HLMS tool in order to introduce the described constraints.

## 7. CIRCUIT SYNTHESIS TOOL (CS)

This tool maps the target complex system on the configurable architecture of the nano-network (figure 6). The input of this tool is the description of the target complex system created by the NAF tool. From this description the CS tool first creates an RTL description of the function executed by the basic module(s) composing the target complex system and of their interconnections. Then it maps this description on the configurable architecture of the nano-network.

The transformation of the high-level description of the function of the basic modules and of their interconnections into an RTL description does not represent a new research challenge as this task is similar to that addressed by high-level synthesis tools, like those described in [17]-[19].

The second component of the synthesis tool is similar to synthesis tools used today to map RTL descriptions on reconfigurable architectures of FPGA families [20]-[22]. Similar approaches would be used for developing a tool that maps RTL descriptions on reconfigurable architectures of nano-networks. This component too does not represent a new challenge.

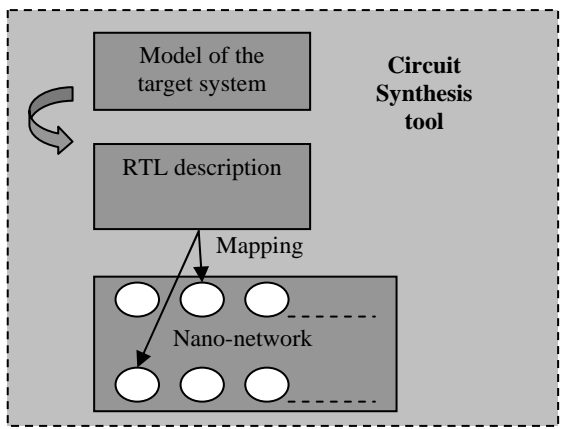

Figure 6: Circuit synthesis tool

## CONCLUSIONS

By considering the constraints for fabricating components including huge amounts of devices (nano-networks), as promised by nanotechnologies, we suggest using a non von Neumann, highly parallel, reconfigurable computing paradigm, for implementing applications of profound interest for science and technology. They consist of complex systems composed of large numbers of simple elements. Then, we describe a CAD platform aimed at exploiting efficiently the nano-networks for implementing this kind of applications. This platform includes:

- A High-level Modeling and Simulation (HLMS) tool allowing implementing in a traditional software approach complex systems composed of a large number of identical modules.
- A Nano-network Architecture FIT (NAF) tool allowing adapting the high-level description generated by HLMS to the nano-network architecture.
- A Circuit Synthesis tool allowing mapping the description generated by the NAF tool on the architecture of the nano-network.

The first of these tools was finalized and described shortly in this paper. It allows generating at low user effort a software representation (model) of the target complex system and exploring its behavior through simulation experiments. The versatility of the tool was illustrated by experimenting two quite different families of complex systems: artificial ecosystems and artificial "universes". Our first experiments of artificial ecosystems showed their rich behavior and the interest to pursue the exploration of more complex cases. The experiments of artificial "universes" illustrated the mechanisms of emergence of relativistic space-time geometry in such systems. Ongoing work concerns the development of the second (NAF) tool.